\documentclass[prl,reprint,amsmath,amssymb,aps,preprintnumbers,showpacs]{revtex4-1}
\usepackage[colorlinks=true,linkcolor=black, citecolor=black,urlcolor=black]{hyperref} 
\usepackage{amstext,amsmath,amssymb,amsfonts,mathrsfs}   
\usepackage{graphicx}  
\usepackage{braket} 
\usepackage{bbm} 
\usepackage{tikz} 
\usetikzlibrary{calc}
\usetikzlibrary{arrows}

\newcommand{\AdS}{\text{AdS}}
\renewcommand{\S}{\text{S}}

\newcommand{\CFT}{\text{CFT}}

\newcommand{\Integers}{\mathbb{Z}}
\newcommand{\Reals}{\mathbb{R}}
\newcommand{\Complex}{\mathbb{C}}

\newcommand{\alg}[1]{\mathfrak{#1}}

\begin{document}

\vspace*{0cm}

\title{The Chern-Simons Origin of Superstring Integrability}
\author{Kevin Costello$^{1}$}
\email{kcostello@perimeterinstitute.ca}
\author{Bogdan Stefa{\'n}ski, jr.$^{2,1}$}
\email{Bogdan.Stefanski.1@city.ac.uk}

\affiliation{$^1$ Perimeter Institute, 31 Caroline St N, Waterloo, Canada\\
${}^2$ Centre for Mathematical Science, City, University of London, Northampton Square, London EC1V 0HB, United Kingdom}

\begin{abstract}
We derive the $\AdS_5\times\S^5$ Green-Schwarz superstring from four-dimensional Beltrami-Chern-Simons theory reduced on a manifold with singular boundary conditions. In this construction, the Lax connection and spectral parameter of the integrable superstring have a simple geometric origin in four dimensions as gauge connection and reduction coordinate. Kappa symmetry arises as a certain class of singular gauge transformations, while the worldsheet metric comes from complex-structure-changing Beltrami differentials. Our approach offers the possibility of investigating integrable holography using traditional field theory methods.
\end{abstract}


\maketitle

\section{Introduction}

Integrability is an invaluable exact tool for $\AdS/\CFT$ holography~\cite{Beisert:2010jr} which has provided significant evidence in favour of many of the dualities conjectured by Maldacena~\cite{Maldacena:1997re}.~\footnote{For a recent pedagogical review of integrability and its holographic applications see~\cite{Bombardelli:2016rwb}.} Signs of integrability were noticed early on in both gauge and string theory. At weak   't Hooft coupling the dilatation operator corresponds to an integrable spin-chain Hamiltonian~\cite{Minahan:2002ve,Beisert:2003yb,Beisert:2003ys}, while at strong coupling superstring equations of motion are equivalent to the flatness of an auxiliary Lax connection~\cite{Bena:2003wd,Mandal:2002fs}. Yet the origin of  the integrable structure underlying these theories remains obscure. 

In this paper we will attempt to demistify the appearance of integrability in holographic superstring theory by showing how Metsaev-Tseytlin (MT) kappa-symmetric string 
actions~\cite{Metsaev:1998it,Henneaux:1984mh} can be obtained from four-dimensional Chern-Simons (CS${}_4$) gauge theory with suitably chosen boundary conditions.  Recently, a new approach to integrability based on CS${}_4$ theory has been proposed in~\cite{Costello:2013zra,Costello:2017dso,Costello:2018gyb}. In this approach, reducing CS${}_4$ theory to two dimensions in the presence of defects~\cite{Costello:2019tri} gives rise to many integrable two-dimensional field theories such as the Gross-Neveu and Wess-Zumino-Witten models, as well as the pure-spinor sigma-model on $\AdS_5\times\S^5$~\cite{Berkovits:2000fe}. 

We begin by obtaining the MT sigma-model from a reduction of CS${}_4$ theory generalising~\cite{Costello:2019tri}. Like in that work, the mysterious Lax connection and spectral parameter of the MT sigma-model have a prosaic interpretation in terms of the gauge potential and direction of reduction in the CS${}_4$ theory~\footnote{This construction generalizes to a large class of MT-like integrable models.}. The absence of fermion kinetic terms makes the MT sigma-model pathological.~\footnote{At quadratic order in fields, the kinetic terms in the MT sigma-model~(9) involve only bosonic fields $J^{(2)}$. This is in counter-distinction to the pure-spinor model which contains additional kinetic terms for fermionic components $J^{(1)}$ and $J^{(3)}$.}  Nonetheless, coupling it to a world-sheet metric leads to a consistent \textit{string theory}~\cite{Arutyunov:2009ga} with  target-space supersymmetry~\cite{Green:1983wt} because the  action has kappa-symmetry~\cite{Siegel:1983hh}.

We couple CS${}_4$ theory to complex-structure-changing Beltrami differentials. In the presence of singular boundary conditions, these couplings cannot be removed by field redefinitions and reducing this Beltrami-Chern-Simons (BCS) theory to two dimensions gives the MT \textit{string theory}. We show that BCS theory has kappa symmetry, with kappa transformations implemented by certain singular \textit{gauge transformations} combined with an action on the Beltrami differentials which has compact support near the singularities. The resulting theory is essentially a conventional gauge theory of Chern-Simons type coupled to extra matter. On general grounds its observables will be Wilson lines whose interactions encode the R matrix of AdS/CFT. These can be computed using Feynman graphs with propagators and vertices derived in a conventional way from the action, taking into account the slightly unusual boundary conditions.  As a result, our construction opens up the possibility of investigating integrability in string theory and holography through conventional and rigorous quantum field theory methods, albeit with singular boundary conditions.

\section{MT sigma-model from CS${}_4$}
\label{sec:MTsm-from-CS4}

Consider the CS${}_4$ action on ${\cal V}= \Sigma\times C$
\begin{equation}
S_{\mbox{\tiny CS}_4}=\frac{1}{h}
\int_{{\cal V}}\,\,\omega\wedge L_{\mbox{\tiny CS}}(A)\,,
\end{equation}
where  $L_{\mbox{\tiny CS}}$ is the CS Lagrangian~\eqref{eq:LCS},
$\Sigma=\Reals^2$, $C=\mathbb{CP}^1$ and the holomorphic one-form $\omega$ has $n$ second-order poles 
 and $2n-2$ first-order zeros 
\begin{equation}
\omega=\frac{\prod (z-q_k)\prod (z-{\tilde q}_k)}{\prod (z-p_i)^2}\mathrm{d} z\,.
\end{equation}
The gauge group of main interest in this paper is $G=\alg{psu}(2,2|4)$ whose dual Coxeter number is zero. As a result, the theory is framing-anomaly free and $h$ is not quantised. As reviewed in~\cite{Arutyunov:2009ga}, $\alg{psu}(2,2|4)$ has a $\Integers_4$ automorphism and we denote its $i^m$ eigenspace by $\alg{psu}(2,2|4)^{(m)}$ ($m=0,\dots,3$) and recall that $\alg{psu}(2,2|4)^{(0)}=\alg{so}(4\,,\,1)\times\alg{so}(5)$. As in~\cite{Costello:2019tri}, for a well-defined action we require $A_{w}\,,A_{\bar{w}}$ to have  first order zeros at $z=p_i$ and  poles at
\begin{equation}
A_w |_{z=q_k}\sim\frac{1}{z-q_k}\,,\qquad\qquad
A_{\bar{w}} |_{z={\tilde q}_k}\sim\frac{1}{z-{\tilde q}_k}\,.
\label{eq:sings}
\end{equation}
Solving the equations of motion and boundary conditions gives $A=\hat{A}+A'$, where $\hat{A}=\hat{\sigma}^{-1}\mathrm{d}\hat{\sigma}$, 
$A'=\hat{\sigma}^{-1}\,{\cal L}\,\hat{\sigma}$, with ${\cal L}_{\bar{z}}=0$ and
\begin{align}
{\cal L}_{w}&=\frac{\prod(z-p_i)}{\prod(z-q_k)}\sum_j\frac{1}{p_j-z}\frac{\prod(p_j-q_k)}{\prod\limits_{i\neq j}(p_j-p_i)}\partial_w\sigma_j\sigma_j^{-1}\,,
\\
{\cal L}_{\bar{w}}&=\frac{\prod(z-p_i)}{\prod(z-{\tilde q}_k)}
\sum_j\frac{1}{p_j-z}\frac{\prod(p_j-{\tilde q}_k)}{\prod\limits_{i\neq j}(p_j-p_i)}\partial_{\bar{w}}\sigma_j\sigma_j^{-1}\,.
\end{align}
The notation used above follows~\cite{Costello:2019tri}. In particular, up to gauge transformations,  $A_{\bar z}$   defines a map $\sigma:\Reals^2\rightarrow G^n$, which extends to  $\hat{\sigma}:{\cal V}\rightarrow G^n$, where near $z= p_i$, $\hat{\sigma}\sim \sigma_i$, with $\sigma_i:\Reals^2\rightarrow G$. In~\cite{Costello:2019tri} constant gauge transformations were used to set $\sigma_n=\mbox{Id}$. Inserting $A$ back into 
$S_{\mbox{\tiny CS}_4}$ localizes the action on the boundary and gives a large family of integrable sigma-models, with left-acting $G^n$ symmetry and Lax connection ${\cal L}$~\cite{Costello:2019tri}, which were also found in~\cite{Delduc:2018hty} without reference to CS${}_4$-theory. 

We now set $p_j=\exp(2\pi i j/n)$  and, in contradistinction to~\cite{Costello:2019tri}, take the limit
\begin{align}
q_1=\dots=q_{n-m}={\tilde q}_{n-m+1}=\dots={\tilde q}_{n-1}
&\rightarrow 0\,,\nonumber \\
{\tilde q}_1=\dots={\tilde q}_{n-m}=q_{n-m+1}=\dots=q_{n-1}
&\rightarrow\infty\,,
\label{eq:GS-limit}
\end{align}
with $1<m<n$. Upon rescaling $h\rightarrow h(-{\tilde q}_1)^{n-1}$, the $m=2$ action is
\begin{align}
S^{n,m=2}&=\frac{k}{4\pi n}\int_\Sigma\sum_i J_{i,w}J_{i,\bar{w}}
-\sum_{i, j}\alpha_{ij}J_{i,\bar{w}}J_{j,w}
\nonumber \\
&\,\,\,
-\frac{k}{8\pi n^2}\int_\Sigma\sum_{i\neq j}
\frac{p_i^3+p_j^3}{p_ip_j(p_i-p_j)}
J_{i,\bar{w}}J_{j,w}
\nonumber \\
&\,\,\,
+\frac{k}{12\pi n^2}
\int_{\Sigma\times \Reals^+}\,f^{abc}X_{i,a}X_{i,b}X_{i,c}\,,
\label{eq:nm-CS-action}
\end{align}
where $J_i\equiv \mathrm{d}\sigma_i\sigma_i^{-1}$,
 $X_i\equiv \sigma_i^{-1}\mathrm{d}\sigma_i$ $f^{abc}$ are the gauge-group structure constants, $\Reals^+$ is the $z$-plane radial direction, $\alpha_{ij}\equiv(1+p_i/p_j+p_j/p_i)/n$ and $k=8\pi i/h$.~\footnote{The factor of $i$ is a consequence of a Euclidean worldsheet.}
 
These models have a $\Integers_n$ symmetry $\rho$, which permutes the $n$ copies of $G$, together with acting by the $\Integers_n$ automorphism on each copy, and multiplies $z$ by an  $n$th root of unity. This fixes a subgroup $\rho(G^{(0)})=G^{(0)}\subset G$, which for $G=\alg{psu}(2,2|4)$ is $G^{(0)}=\alg{so}(4,1)\times\alg{so}(5)$. Gauging the $\Integers_n$ action leads to novel integrable models on generalised symmetric spaces $G/G^{(0)}$ whose equations of motion are equivalent to the flatness condition of the Lax connection
\begin{align}
{\cal L}^{n,m}_w(z)
&= \sum_{k=0}^m z^k J^{(k)}_w + \sum_{k=m+1}^{n-1} z^{k-n} J^{(k)}_w\,,
\nonumber\\
{\cal L}^{n,m}_{\bar{w}}(z)
&= \sum_{k=0}^{n-m-1} z^k J^{(k)}_{\bar{w}} + \sum_{k=n-m}^{n-1} z^{k-n} J^{(k)}_{\bar{w}}\,,
\end{align}
where $J_a\equiv J_{1,a}$, $J_{i,a}=\rho^{i-1}(J_a)$ and $J^{(k)}$ is the $k$-th $\Integers_n$-eigenspace. In particular, for $m=2$, $n=4$ we obtain the MT $\sigma$-model~\cite{Metsaev:1998it,Henneaux:1984mh,Berkovits:1999zq} with target-space $\alg{psu}(2,2|4)/(\alg{so}(4,1)\times\alg{so}(5))$
\begin{equation}
S^{n=4,m=2}=\frac{k}{4\pi}\int_\Sigma\,J^{(2)}_wJ^{(2)}_{\bar{w}}
-J^{(1)}_wJ^{(3)}_{\bar{w}}+J^{(3)}_wJ^{(1)}_{\bar{w}}\,.
\label{eq:MT-sigma}
\end{equation}

\section{Metric and Virasoro constraints}
\label{sec:metric}

The $S^{n,m}$ models, unlike their pure-spinor counterparts~\cite{Costello:2019tri}, do not have second-order kinetic terms for fermions, since the boundary conditions imposed in CS${}_4$ to obtain them are not elliptic. This apparent disadvantage is in fact a boon:
for judicious choices of $G$, $S^{4,2}$ has kappa-symmetry when coupled to the worldsheet metric. 

In our derivation of $S^{4,2}$ from CS${}_4$-theory it is not immediately clear how a worldsheet metric might arise. Afterall, CS${}_4$-theory does not depend on the four-dimensional metric. More precisely, this is only true for manifolds without boundary or for everywhere regular field configurations. On the other hand, allowing singularities in the gauge field like those in equation~\eqref{eq:sings} means that the CS${}_4$-theory may no longer be invariant under general coordinate transformations on the boundary or in regions near such singularities. If we separate the singular part of the gauge connection
\begin{equation}
A_w\equiv A_w^{(\mathrm{p})} + A_w^{(\mathrm{reg})}\,,\qquad
A_{{\bar w}}\equiv A_{{\bar w}}^{(\mathrm{p})} + A_{{\bar w}}^{(\mathrm{reg})}\,,
\end{equation}
$S_{\mbox{\tiny CS}_4}$ is not  general-coordinate invariant
\begin{equation}
\delta_v S_{\mbox{\tiny CS}_4}\sim -\!\!\!\!\oint\displaylimits_{z=0\,,\infty} \!\!\!\omega\,{}_\wedge\left(
A_w^{(\mathrm{p})} \mathrm{d}v^w
+A_{{\bar w}}^{(\mathrm{p})} \mathrm{d}v^{{\bar w}}
\right){}_\wedge\, A\,.
\end{equation}
We  used Green's theorem to re-write $\delta_v S_{\mbox{\tiny CS}_4}$ as a contour integral around  $z=0\,,\infty$, where $A_w$ and $A_{{\bar w}}$ have poles. For $S^{4,2}$ this integral reduces to
\begin{equation}
 -\int_\Sigma\left(
A_w^{(\mathrm{2})} A_w^{(\mathrm{2})} \mathrm{d}w{}_\wedge \mathrm{d}v^w
+
A_{{\bar w}}^{(\mathrm{2})} A_{{\bar w}}^{(\mathrm{2})} \mathrm{d}{\bar w}{}_\wedge \mathrm{d}v^{\bar w}\right)\,.
\label{eq:diff-fail}
\end{equation}
Since this is non-zero, it appears that with our boundary conditions the CS${}_4$-theory is no longer invariant under general coordinate transformations.

The lack of diffeomorphism invariance~\eqref{eq:diff-fail} suggests we need to introduce a new field to restore it. We do this by varying the complex structure with the new field corresponding to the Beltrami differential $\beta$.
Under such a variation, the de Rham differential $\mathrm{d}$, which is 
the  CS${}_4$-theory's BRST operator, changes to
\begin{equation}
\mathrm{d}\longrightarrow \mathrm{d}
+\mathrm{d}\bar{w}\,{\cal L}_{\beta_{\bar{w}}\partial_w}
+\mathrm{d}\bar{z}\,{\cal L}_{\beta_{\bar{z}}\partial_w}
\,,\label{eq:beltr-coupl}
\end{equation}
where ${\cal L}$ is the Lie derivative. Since the Cartan homotopy formula gives the Lie derivative as
\begin{equation}
{\cal L}_{V} = \left[\mathrm{d}\,,\,\iota_{V}\right]\,,
\end{equation}
where $\iota_V$ is the inner multiplication (contraction) by $V$, we are adding a BRST-exact term to the action, which should have no effect in the bulk.  
The $\beta_{\bar{w}}$-dependent part of the action is
\begin{equation}
S_{\beta_{\bar{w}}}\equiv\frac{2}{h}\int_{{\cal V}}\,\,\omega_z \beta_{\bar{w}}
A_w(\partial_{\bar{z}} A_w-\partial_w A_{\bar{z}})\,,
\label{eq:belt-cs-terms}
\end{equation}
with a similar term for $S_{\beta_{\bar{z}}}$. The combined Beltrami-Chern-Simons action 
\begin{equation}
S_{\mbox{\tiny BCS}}\equiv
S_{\mbox{\tiny 4d}}+S_{\beta_{\bar{w}}}+S_{\beta_{\bar{z}}}
\end{equation}
is invariant, up to a $w$-derivative, under a new gauge invariance
\begin{align}
A\longrightarrow A+{\cal L}_V A\,,\qquad
\beta\longrightarrow \beta+{\cal L}_V \beta\,,
\label{eq:beta-gauge-inv}
\end{align}
with gauge parameter $V\equiv v\,\partial_w$ for an arbitrary function $v$. This invariance can be used  to gauge away one of the components of $\beta$, for example $\beta_{\bar{z}}\rightarrow 0$. Working in this gauge, redefining $A_{\bar w}$ 
\begin{equation}
A_{\bar{w}}\longrightarrow A_{\bar{w}}-\beta_{\bar{w}}A_w\,,
\label{eq:field-redef}
\end{equation}
one recovers the original CS${}_4$ action
\begin{align}
S_{\mbox{\tiny CS}_4}+S_{\beta_{\bar{w}}}\longrightarrow &\,\, S_{\mbox{\tiny CS}_4}
\end{align}
This is to be expected, since on a manifold with no boundary the CS${}_4$-theory is metric-independent and we are adding a BRST-exact Beltrami term to it. This should leave the theory unmodified, up to field redefinitions. However, in the presence of a boundary the field redefinition~\eqref{eq:field-redef} might not be compatible with the boundary conditions. Indeed, $A_{\bar{w}}$ and $A_w$ have poles of different order at $z=0\,,\,\infty$, while $A_{\bar{z}}$ is regular. Since $\beta$ should be regular on the boundary, this means that we cannot eliminate the Beltrami couplings on the boundary using field redefinitions~\eqref{eq:field-redef}.
With boundary conditions~\eqref{eq:GS-limit} and~\eqref{eq:sings}, the $\beta$-dependent part of the action reduces to a boundary contribution at $z=0$
\begin{equation}
S_\beta=\,\frac{ \delta_{z=0}}{h}\int_\Sigma \, 2\beta_{\bar{w}}\,A^{(2)}_w  A^{(2)}_w\,.
\label{eq:met-gt}
\end{equation}
This coupling restores diffeomorphism invariance at $z=0$ and varying the action with respect to $\beta_{\bar{w}}$ leads to the Virasoro constraint 
\begin{equation}
A_w^{(\mathrm{2})} A_w^{(\mathrm{2})}=0\,.
\label{eq:vir-const}
\end{equation}
We can introduce a similar modification to equation~\eqref{eq:beltr-coupl} along 
$\partial_{\bar{w}}$ and show that it is trivial up to a field redefinition away from $z=\infty$, leading to a boundary action
\begin{equation}
S_{{\tilde \beta}_w}=\,\frac{\delta_{z=\infty}}{h} \int_\Sigma \, 2 {\tilde \beta}_{w}\,A^{(2)}_{\bar{w}}  A^{(2)}_{\bar{w}}\,,
\label{eq:met-gt2}
\end{equation}
which restores diffeomorphism invariance at $z=\infty$.

In the Polyakov action, the world-sheet metric $g$ is taken up to Weyl transformations. If we analytically continue this action to allow $g$ to be a metric with complex coefficients, then the data of $g$, up to Weyl transformations, is equivalent to the data of a holomorphic Beltrami differential $\beta$ and an anti-holomorphic Beltrami differential $\bar{\beta}$~\footnote{To see this, recall that a conformal structure on an oriented surface is the same data as a complex structure, or equivalently, an anti-complex structure. If we analytically continue, these two piece of data need no longer be the same, so that an analytically-continued conformal deformation of the Riemann surface $\Sigma$ is equivalent to a deformation of the complex structure $\Sigma$ and of the complex conjugate surface $\bar{\Sigma}$. In more prosaic terms, a deformation of the metric $g$ has $3$ components, one of which can be removed by a Weyl transformation. The other two are, once we complexify, sections of $(T^{1,0}\Sigma)^{\otimes 2}) $ and $(T^{0,1}\Sigma)^{\otimes 2}) $. Using the metric we are deforming to identify $T^{0,1}\Sigma$ with the dual of $T^{1,0}\Sigma$, these two components become the holomorphic and anti-holomorphic Beltrami differential.}. The reality condition corresponding to asking that $g$ be a metric with real coefficients is that we ask $\beta$ and $\bar{\beta}$ to be complex conjugate. For a discussion of such a factorization of the Polyakov action see~\cite{Gwilliam:2017axm}. 

Similarly, in the Polyakov action, the gauge symmetries (after gauging away Weyl transformations) are world-sheet diffeomorphisms. Infinitesimally these are sections of the tangent bundle $T \Sigma$. If we analytically continue the Polyakov action, allowing the infinitesimal world-sheet diffeomorphisms to be complex, we find the gauge transformations we used for the Beltrami differential fields. Indeed, $T\Sigma \otimes_{\Reals}\Complex$ decomposes as $T^{1,0}\Sigma$, which gives the gauge transformations for $\beta$, and $T^{0,1}\Sigma$, giving the gauge transformations for $\bar{\beta}$.

\section{Gauge invariance in BCS theory}

Before proceeding to discuss kappa symmetry in BCS theory, we briefly review how to modify gauge variations in the presence of a Beltrami deformation $\beta_w$ in order for $S_{\mbox{\scriptsize BCS}}$ to be gauge invariant. In components the CS Lagrangian is
\begin{equation}
\tfrac{1}{2}L_{\mbox{\tiny CS}}=
A_{\bar{w}}\left(\partial_{\bar{z}}A_{w}-\partial_{w}A_{\bar{z}}\right)
+A_{\bar{z}}\partial_w A_{\bar{w}}
-A_{\bar{w}}\left[A_w,A_{\bar{z}}\right]\,,
\label{eq:LCS}
\end{equation}
where 
as in the rest of the paper the trace is implicit. The $\beta_w\equiv\beta$ part of the Beltrami-deformed action is
\begin{equation}
S_\beta=-\frac{1}{h}
\int_{\cal V}
\omega_z\left(
\partial_{\bar{z}}\beta_{\bar{w}} \,A_w^2+2\beta_{\bar{w}}\,A_w\,\partial_wA_{\bar{z}}\right)\,.
\end{equation}
$S_{\mbox{\scriptsize CS}_4}$ is invariant under gauge transformations
\begin{equation}
\delta_\chi A_\mu=\partial_\mu\chi+\left[A_\mu\,,\,\chi\right]\,,
\end{equation}
but $S_\beta$ is not
\begin{align}
\delta_\chi S_\beta
&=2
\int_{\cal V} \omega\,\beta_{\bar{w}}\bigl( \partial_{\bar{z}}A_w
-\partial_w A_{\bar{z}}- \left[A_w , A_{\bar{z}}\right]\bigr)\partial_w\chi
\,.
\label{eq:gauge-var-Sb}
\end{align}
To cancel this we modify the gauge-variation of $A_{\bar{w}}$ in accordance with equation~\eqref{eq:beltr-coupl} to
\begin{equation}
\delta_{\chi,\beta} \,A_{\bar{w}}=\partial_{\bar{w}}\chi+\left[A_{\bar{w}}\,,\,\chi\right]-\beta_{\bar{w}}\partial_w\chi\,,
\end{equation}
while leaving the gauge variation of the other components of $A$ unchanged. Since $S_\beta$ does not depend on $A_{\bar{w}}$ its variation~\eqref{eq:gauge-var-Sb} is unchanged, while the gauge variation of $S_{\mbox{\scriptsize 4d}}$ becomes
\begin{align}
\delta_{\chi,\beta}S_{\mbox{\scriptsize CS}_4}
&=-\delta_\chi S_\beta
\,,
\end{align}
making $S_{\mbox{\scriptsize BCS}}$ gauge-invariant.

\section{Kappa symmetry}
\label{sec:kappa}

\noindent We now show that the action $S_{\mbox{\tiny BCS}}$ is invariant under certain singular $G$-gauge transformations, which reduce to kappa-symmetry in the $\sigma$-model. To this end, consider gauge variations 
\begin{equation}
\delta_\xi A =	\mathrm{d} \xi + \left[ A\,,\,\xi\right]
\end{equation}
which have a simple pole near $z=0$
\begin{equation}
\label{eq:g-var-z0-pole}
\xi \sim \frac{1}{z} \xi^{(3)}+\dots \,.
\end{equation}
Since the generator of the $\Integers_4$ automorphism multiplies $z$ by $-i$, we immediately see that the singular gauge variation $\xi^{(3)}$ must be in the $i^3$ eigenspace of $\Integers_4$ and hence is fermionic, as expected of a kappa-variation. The variation of $S_{\mbox{\tiny CS}_4}$ is
\begin{equation}
\delta_\xi S_{\mbox{\tiny CS}_4}
=\frac{1}{h}\int_{\cal V}
\omega\, \left( A_{{\bar w}} \mathrm{d}_{{\bar z}}  \left[A_w \,,\, \xi \right]  - A_w \mathrm{d}_{{\bar z}}  \left[A_{{\bar w}} \,,\, \xi \right] \right)\,.
\label{eq:bdry-var-cs}
\end{equation}
Near $z=0$ the gauge fields have an expansion
\begin{equation}
A_{{\bar w}} \sim \frac{A^{(3)}_{{\bar w}}}{z} +\dots\,, \qquad
 A_w        \sim \frac{A^{(2)}_w}{z^2} + \frac{A^{(3)}_w}{z}+\dots\,.
\label{eq:z-is-0-exp}
\end{equation}
Inserting these into~\eqref{eq:bdry-var-cs} we get~\footnote{We use the fact that  $\mathrm{tr}\left(g^{(3)}h^{(2)}\right)=0$ for arbitrary elements $g$, $h$ of the Lie algebra. 
}
\begin{align}
\delta_\xi S_{\mbox{\tiny CS}_4} 
&=\,\,\frac{\delta_{z=0}}{h}\int_\Sigma  \,
\left[A^{(3)}_{{\bar w}}\,,\, A^{(2)}_w\right]\xi^{(3)}
\,.
\end{align}
 Analogously, for gauge variations with a simple pole at $z=\infty$, we have $\xi \sim z \xi^{(1)}+\dots$,   and the CS${}_4$ action changes by
\begin{equation}
\delta_{\tilde\xi} S_{\mbox{\tiny CS}_4}  = \,\,\frac{\delta_{z=\infty}}{h}\int_\Sigma 
\left[A^{(1)}_{w}\,,\, A^{(2)}_{{\bar w}}\right]{\tilde\xi}^{(1)}\,.
\end{equation}
Kappa transformations can be obtained from the singular gauge transformations by requiring~\cite{Arutyunov:2009ga}
\begin{align}
{\tilde\xi}^{(1)}&\equiv A^{(2)}_{{\bar w}} \kappa^{(1)}_{w}+ \kappa^{(1)}_{w} A^{(2)}_{{\bar w}} \,,
\nonumber \\
\xi^{(3)}&\equiv A^{(2)}_{w} \kappa^{(3)}_{{\bar w}}+ \kappa^{(3)}_{{\bar w}} A^{(2)}_{w} \,,
\end{align}
where the $\kappa$ are the independent (local) parameters of kappa transformations. Notice that the above expression involves selecting a particular (matrix) representation for the gauge group, and using matrix multiplication in that representation. In judiciously chosen cases, there are certain famous Fierz identities~\cite{Brink:1976bc,Gliozzi:1976qd,Arutyunov:2009ga} (see equation (1.80) of~\cite{Arutyunov:2009ga}) that can be used to re-express the kappa variation of $S_{\mbox{\tiny CS}_4}$ as
\begin{align}
\delta_\kappa S_{\mbox{\tiny CS}_4}
 =&\,-\frac{\delta_{z=0}}{2h} \int_\Sigma  
\mathrm{tr}\bigl(A^{(2)}_{w}A^{(2)}_{w}\bigr)
\mathrm{tr}\left(\!\Upsilon [\kappa^{(3)}_{{\bar w}},A^{(3)}_{{\bar w}}]\right)
\nonumber \\
&-\frac{\delta_{z=\infty}}{2h}\int_\Sigma  
\mathrm{tr}\bigl(A^{(2)}_{{\bar w}}A^{(2)}_{{\bar w}}\bigr)
\mathrm{tr}\left(\!\Upsilon [\kappa^{(1)}_{w},A^{(1)}_{w}]\right)
\,.
\label{eq:non-g-inv-sing}
\end{align}
Above, $\Upsilon$ is a suitable constant matrix which for $\alg{psu}(2,2|4)$ is $\mbox{diag}(1_4,-1_4)$. The lack of gauge invariance under~\eqref{eq:g-var-z0-pole} can be compensated  by varying the Beltrami operators under kappa transformations.  Working in the gauge $\beta_{\bar{z}}=0$, we demand 
\begin{equation}
\delta_\kappa\beta_{\bar{w}}=\frac{\delta_{|z|\le \varepsilon}}{2}\mathrm{tr}\left(\!\Upsilon [\kappa^{(3)}_{{\bar w}},A^{(3)}_{{\bar w}}]\right)\,,
\end{equation}
where $\delta_{|z|\le \varepsilon}$ has support in an $\varepsilon$-neighbourhood of $z=0$ only.
Now, $S_{\beta_{\bar{w}}}$ is no longer gauge invariant near $z=0$. Expanding as in equation~\eqref{eq:z-is-0-exp}, we find 
\begin{align}
\delta_\kappa S_\beta &=\, \frac{1}{2h}\int_{{\cal V}_4}
 \frac{\partial_{\bar{z}}\delta_{|z|\le \varepsilon}}{z}\mathrm{tr}\left(\!\Upsilon [\kappa^{(3)}_{{\bar w}},A^{(3)}_{{\bar w}}]\right)
\mathrm{tr}\bigl(A^{(2)}_wA^{(2)}_w\bigr)
\nonumber \\
&=\frac{1}{2h} \oint\displaylimits_{z=0} \frac{\mathrm{d}z}{z}\int_\Sigma\,\mathrm{tr}\left(\!\Upsilon [\kappa^{(3)}_{{\bar w}},A^{(3)}_{{\bar w}}]\right)
\mathrm{tr}\bigl(A^{(2)}_wA^{(2)}_w\bigr)
\end{align}
using the identity $\partial_{\bar{z}}\delta_{|z|\le \varepsilon}=\delta_{|z|= \varepsilon}$.
This cancels the gauge non-invariance of  $S_{\mbox{\tiny CS}_4}$ at $z=0$ in equation~\eqref{eq:non-g-inv-sing}. The $z=\infty$ term can be analogously canceled by a
${\tilde\beta}_w$ variation. The seperate cancelations at $z=0,\infty$, which on the worldsheet correspond to self-dual- and anti-self-dual-vector representations or after Wick rotation holomorphic and anti-holomorphic ones, provide a novel separation of the two sectors in four dimensions.

\section{Conclusions}

In this paper we have introduced  BCS theory and showed that, upon imposing suitable singular boundary conditions, it reduces to the MT superstring. The Beltrami fields and boundary gauge connection of  BCS theory map to the world-sheet metric and sigma-model fields, respectively. The Lax connection and spectral parameter  appear somewhat mysteriously in string theory, but from the four-dimensional BCS point of view they are simply the gauge connection and holomorphic coordinate of the reduction. In BCS theory kappa-symmetry corresponds to a certain class of singular gauge transformations and kappa-invariance holds for gauge groups for which a suitable hypercharge matrix $\Upsilon$ exists. This includes $G=\alg{psu}(2,2|4)$ and its plane-wave~\cite{Metsaev:2002re} and flat-space~\cite{Henneaux:1984mh} limits. We will investigate the properties of these backgrounds from the four-dimensional point of view more fully in a forthcoming paper~\cite{future}.  

There are other well-known integrable superstring backgrounds with a Lax 
connection~\cite{Stefanski:2008ik,Arutyunov:2008if,Babichenko:2009dk,OhlssonSax:2011ms,Cagnazzo:2012se,Sorokin:2011rr,Hoare:2014kma}. In these cases, the MT coset action often needs to be suplemented by extra fermionic degrees of freedom~\cite{Gomis:2008jt} to obtain an action equivalent to the conventional kappa-symmetric superstring actions~\cite{Grisaru:1985fv,Duff:1987bx} and it would be interesting to see how to extend these to  BCS theory. Some of these  backgrounds have target-space moduli and understanding how these appear in BCS theory could provide new insights into moduli spaces.

It would be interesting to perform a Batalin–Vilkovisky quantisation of BCS theory. This introduces a tower of extra fields incuding conventional $b-c$ ghosts of string theory. If the $\Reals^2$ with coordinates $w\,,{\bar w}$ is replaced by a Riemann surface $\Sigma$, then the formalism we have described includes the integral over the moduli of the world-sheet $\Sigma$. Indeed, the Beltrami differential $\beta$ on $\Sigma$ has zero-modes which live in the Dolbeault cohomology group $H^1(\Sigma, T\Sigma)$, which is of (complex) dimension $3g-3$ for $g>1$.~\footnote{For g=1 the Beltrami differential has one zero mode.} The anti-holomorphic Beltrami differential has zero-modes which live in the complex conjugate of this space. Together, the manifold of zero-modes is the product of the moduli space with its complex conjugate: $\mathcal{M}_g \times \bar{\mathcal{M}}_g$. As we are doing an analytically-continued path integral, we need to choose an integration contour. It is natural to choose the contour to be the locus where the holomorphic and anti-holomoprhic Beltrami differential are complex conjugate, leading to an integral over one copy of the moduli space $\mathcal{M}_g$. We will return to a detailed discussion of this and its relation to the Polyakov path integral over Riemann surfaces~\cite{Polyakov:1981rd} in a future paper~\cite{future}. 

We hope our construction can shed light on the relationship between the pure-spinor and Green-Schwarz formulations of string theory. Quantising BCS theory should also offer new insights into quantum integrability of holographic string backgrounds and connect with the Quantum Spectral Curve approach~\cite{Gromov:2013pga,Bombardelli:2017vhk}. In particular, we expect that compactifying the string  worldsheet $\Sigma=\Reals\times\mbox{S}^1$ will lead to diagrams involving the photon propagator `wrapping' the $\mbox{S}^1$ direction and the $\mathbb{Q}$-functions should appear as solutions of the Baxter equation involving the transfer matrix of BCS theory.

\bigskip
We would like to thank Masahito Yamazaki for discussions. This research was supported in part by a grant from the Krembil Foundation. K.C. is supported by the NSERC Discovery Grant program and by the Perimeter Institute for Theoretical Physics. BS acknowledges funding support from an STFC Consolidated Grant “Theoretical Physics at City University” ST/P000797/1. BS is grateful for the hospitality of Perimeter Institute where part of this work was carried out. Research at Perimeter Institute is supported in part by the Government of Canada through the Department of Innovation, Science and Economic Development Canada and by the Province of Ontario through the Ministry of Colleges and Universities. 
\\

\bibliographystyle{h-physrev}
\bibliography{GSCS-short-v3}

\end{document}